\documentclass[apj,twocolappendix,numberedappendix]{emulateapj}
\pdfoutput=1
\usepackage{hyperref}
\usepackage{graphicx}
\usepackage{times}
\usepackage{amsmath}
\usepackage{color}

\DeclareGraphicsExtensions{.pdf}

\shortauthors{\sc Clark et al.}

\begin{document}

\title{PSR~J1906$+$0722: An Elusive Gamma-Ray Pulsar} 

\author{
C.~J.~Clark\altaffilmark{1,2},
H.~J.~Pletsch\altaffilmark{1,2},
J.~Wu\altaffilmark{3},
L.~Guillemot\altaffilmark{4,5},
M.~Ackermann\altaffilmark{6},
B.~Allen\altaffilmark{1,7,2},
A.~de~Angelis\altaffilmark{8},
C.~Aulbert\altaffilmark{1,2},
L.~Baldini\altaffilmark{9,10},
J.~Ballet\altaffilmark{11}
% ...
}
\altaffiltext{1}{Albert-Einstein-Institut, Max-Planck-Institut f\"ur Gravitationsphysik, D-30167
  Hannover, Germany; \href{mailto:colin.clark@aei.mpg.de}{colin.clark@aei.mpg.de}}
\altaffiltext{2}{Leibniz Universit\"at Hannover, D-30167 Hannover, Germany}
\altaffiltext{3}{Max-Planck-Institut f\"ur Radioastronomie, Auf dem H\"ugel 69, D-53121 Bonn, Germany}
\altaffiltext{4}{Laboratoire de Physique et Chimie de l'Environnement et de l'Espace -- Universit\'e d'Orl\'eans / CNRS, F-45071 Orl\'eans Cedex 02, France}
\altaffiltext{5}{Station de radioastronomie de Nan\c{c}ay, Observatoire de Paris, CNRS/INSU, F-18330 Nan\c{c}ay, France}
\altaffiltext{6}{Deutsches Elektronen Synchrotron DESY, D-15738 Zeuthen, Germany}
\altaffiltext{7}{Department of Physics, University of Wisconsin-Milwaukee, P.O. Box 413, Milwaukee, WI 53201, USA}
\altaffiltext{8}{Dipartimento di Fisica, Universit\`a di Udine and Istituto Nazionale di Fisica Nucleare, Sezione di Trieste, Gruppo Collegato di Udine, I-33100 Udine}
\altaffiltext{9}{Universit\`a di Pisa and Istituto Nazionale di Fisica Nucleare, Sezione di Pisa I-56127 Pisa, Italy}
\altaffiltext{10}{W. W. Hansen Experimental Physics Laboratory, Kavli Institute for Particle Astrophysics and Cosmology, Department of Physics and SLAC National Accelerator Laboratory, Stanford University, Stanford, CA 94305, USA}
\altaffiltext{11}{Laboratoire AIM, CEA-IRFU/CNRS/Universit\'e Paris Diderot, Service d'Astrophysique, CEA Saclay, F-91191 Gif sur Yvette, France}
% ...

\begin{abstract} 
\noindent
We report the discovery of PSR~J1906$+$0722, a gamma-ray pulsar detected as part of a blind survey of unidentified \textit{Fermi} Large Area Telescope (LAT) sources being carried out on the volunteer distributed computing system, \textit{Einstein@Home}. 
This newly discovered pulsar previously appeared as the most significant remaining unidentified
gamma-ray source without a known association in the second \textit{Fermi}-LAT source catalog (2FGL) and was among the top ten most significant unassociated sources in the recent third catalog (3FGL).
PSR~J1906$+$0722 is a young, energetic, isolated pulsar, with a spin frequency of $8.9$ Hz, a characteristic age of $49$\,kyr, and spin-down power $1.0 \times 10^{36}$\,erg\,s$^{-1}$. In 2009~August it suffered one of the largest glitches detected from a gamma-ray pulsar ($\Delta f / f \approx 4.5\times10^{-6}$).
Remaining undetected in dedicated radio follow-up observations, the pulsar is likely radio-quiet.  An off-pulse analysis of the gamma-ray flux from the location of PSR~J1906$+$0722 revealed the presence of an additional nearby source, which may be emission from the interaction between a neighboring supernova remnant and a molecular cloud.
We discuss possible effects which may have hindered the detection of PSR~J1906$+$0722 in previous searches and describe the methods by which these effects were mitigated in this survey. We also demonstrate the use of advanced timing methods for estimating the positional, spin and glitch parameters of difficult-to-time pulsars such as this.
\end{abstract} 


\section{Introduction}\label{s:intro}
The large collecting area and continuous observation mode of the \textit{Fermi} Large Area Telescope \citep[LAT;][]{generalfermilatref} make it an ideal instrument for the detection and analysis of periodic gamma-ray emission from pulsars. Through the careful analysis of the arrival times of photons covering the $6$ years since its launch, the LAT has discovered pulsed gamma-ray emission from more than 160 pulsars\footnote{\url{http://tinyurl.com/fermipulsars}} \citep{2PC+2013,Caraveo2014}.

While the majority of these pulsars were first found in radio observations \citep[e.g.,][]{Abdo2009+J1028,Abdo2009+J0030}, the ephemerides from which could be used to test for gamma-ray pulsations,
a substantial fraction of the gamma-ray pulsar population was discovered through blind searches of \textit{Fermi}-LAT data \citep[e.g.,][]{Abdo2009+Blind16,SazParkinson2010+Blind8}.

In a recent work \citep{Methods2014} we presented newly advanced methods  designed to increase the sensitivity of blind searches without increasing the computational cost. These improvements have since been incorporated into a new blind survey of unidentified, pulsar-like \textit{Fermi}-LAT sources being conducted on the distributed volunteer computing system, \textit{Einstein@Home}.\footnote{\url{http://www.einsteinathome.org}} Previous surveys have been extremely successful in detecting new gamma-ray pulsars \citep{Pletsch+2012-9pulsars,Pletsch2012+J1838,Pletsch+2013-4pulsars}, and the newly improved search methods, in combination with the latest \textit{Fermi}-LAT data, offer a significant increase in sensitivity. 

As part of this survey, we carried out a blind search for pulsed emission from a point source in the third \textit{Fermi}-LAT source catalog \citep[3FGL,][]{3FGL}, 3FGL~J1906.6$+$0720. 
This source, previously known as 2FGL~J1906.5$+$0720 \citep{2FGL}, is highly significant and stands out as the most significant unassociated 2FGL source. Moreover, it was included in the ``bright'' pulsar-like source list described by \cite{Romani2012}.
An investigation of the spectral properties of 2FGL sources found that, after the source associated with the Galactic Center, 2FGL~J1906.5$+$0720 was the unidentified source most likely to contain a pulsar \citep{Lee2012+GMM}. As such, over recent years, this source has been searched for pulsations, both in gamma rays \citep[e.g.,][]{Pletsch+2012-9pulsars,Xing2014} and in radio observations \citep[e.g.,][]{Barr2013}.
However, despite these attempts, pulsed emission from this source remained undetected until now.

Here, we present the discovery and follow-up study of PSR~J1906$+$0722, a young isolated gamma-ray pulsar detected by the \textit{Einstein@Home} survey.

\section{Discovery}\label{s:discovery}

\subsection{Data Preparation}
In the blind search we analyzed \emph{Fermi}-LAT data recorded 
between 2008 August 4 and 2014 April 6. The \emph{Fermi} Science Tools\footnote{\url{http://fermi.gsfc.nasa.gov/ssc/data/analysis/software}} were used to extract Pass 8 source class
photons, which were analyzed using the \texttt{P8\_SOURCE\_V3} instrument response functions (IRFs).\footnote{The Science Tools, IRFs and diffuse models used here are internal pre-release versions of the Pass 8 data analysis. Our results did not change substantially with the final release versions.}
We used \texttt{gtselect} to select photons with reconstructed directions within an 8$\arcdeg$ region of interest (ROI) around
3FGL~J1906.6$+$0720, photon energies $>100\,$MeV and zenith angles $< 100\arcdeg$.
We only included photons detected when the LAT was working in normal science mode, and with rocking angle $< 52\arcdeg$. 

To assign photon weights representing the probability of each photon having been emitted by the target source \citep{Kerr2011}, we performed a likelihood spectral analysis using the \texttt{pointlike} package. 
We built a source model by including
all 3FGL catalog sources located within 13$\arcdeg$ of 3FGL~J1906.6$+$0720, while allowing the spectral parameters of point sources within 5$\arcdeg$ to vary. We modeled the gamma-ray spectrum 
of this source with an exponentially cutoff power law, typical of gamma-ray 
pulsar spectra \citep{2FGL}. We used the \texttt{template\_4years\_P8\_V2\_scaled.fits} map 
cube and \texttt{isotropic\_source\_4years\_P8V3} template to model the Galactic diffuse emission and the isotropic
diffuse background respectively.\footnote{\url{http://fermi.gsfc.nasa.gov/ssc/data/access/lat/BackgroundModels.html}} The normalization parameters of both diffuse components were left free. Finally, the photon weights were computed using \texttt{gtsrcprob}, based on the
best-fit source model resulting from the likelihood analysis. 

\subsection{Blind Search Method}\label{s:blind_search}
For the blind search, we assumed a canonical isolated pulsar model, making it necessary to search in
four parameters: spin frequency, $f$, spin-down rate, $\dot{f}$, R.A., $\alpha$ and decl., $\delta$.

The basis for most blind searches for gamma-ray pulsars is the well-known multistage scheme based around an initial semicoherent search \citep[e.g.,][]{Atwood2006,Pletsch+2012-9pulsars}. For this survey, we implemented the form of the multistage search scheme described in \cite{Methods2014}, where the initial semicoherent stage uses a lag-window of duration $2^{21}$s $\approx 24$ days. 

Notably, this survey incorporates an intermediate semicoherent refinement step, with a longer (more sensitive)  lag-window of $2^{22}$s $\approx 48$ days, reducing the parameter space around each first-stage candidate to be searched in the final fully-coherent follow-up step. This improves the efficiency of the follow-up stage, and allows the search to ``walk'' away (in all 4 search parameters) from the original location of the candidate if necessary.

\begin{figure}
	\centering
	\includegraphics[width=\columnwidth]{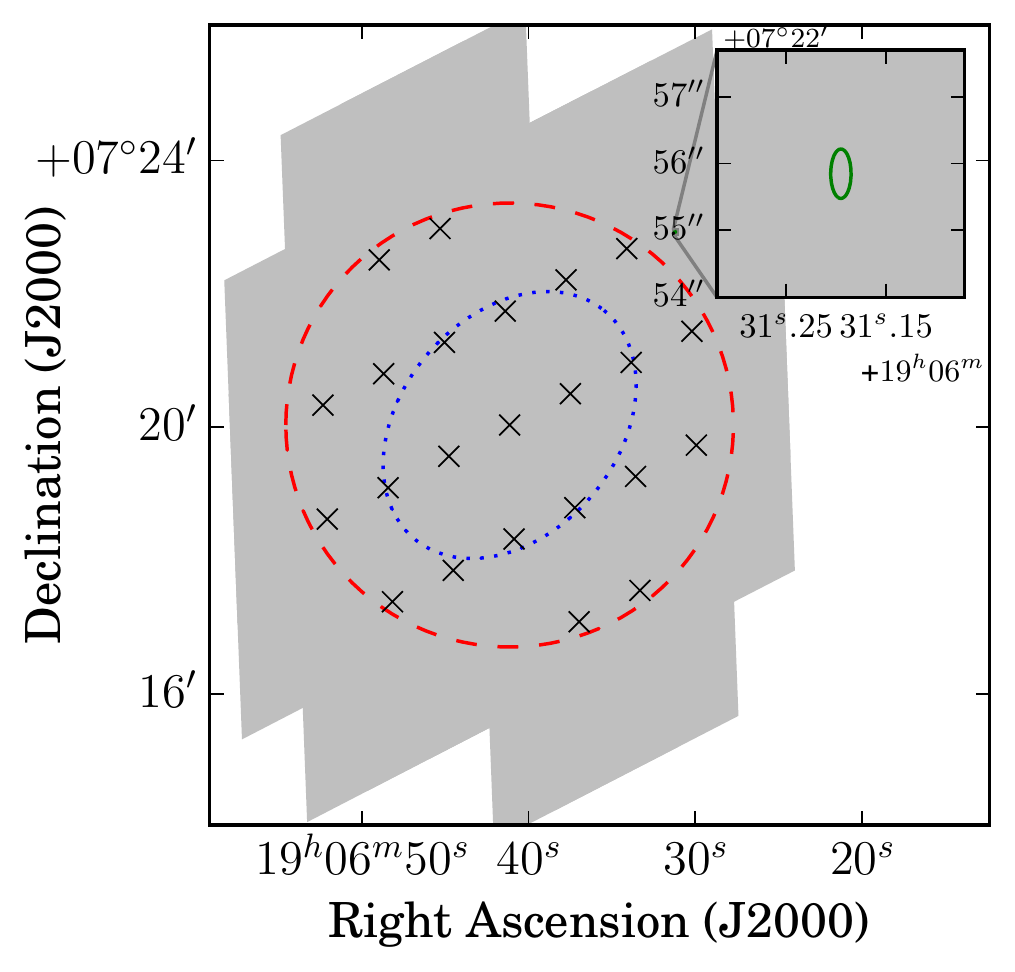}
	\caption{Sky location of PSR~J1906$+$0722 and positional offset from the catalog location. The dotted ellipse shows the 3FGL 95\% confidence region. The dashed ellipse shows the region in which the search grid (crosses) was constructed for the initial semicoherent search stage. The filled area shows the region that can be reached by the follow-up stage as it moves away from the initial candidate location. The $1\sigma$ region from the timing solution, shown by the solid ellipse, is highlighted in the inset.}
	\label{f:sky_regions}
\end{figure}

Figure~\ref{f:sky_regions} illustrates the importance of these new techniques. 
In the blind survey, we searched a conservatively large circular region around the 3FGL sky location with a radius $50$\% larger than the 3FGL $95$\% confidence region.
As evident from Figure~\ref{f:sky_regions}, the pulsar lies far outside the original source's confidence region, and also outside our search region. We therefore owe its detection to the large resolution of the semicoherent step, and the flexibility of the follow-up steps, which allow for signals to be detected despite a large offset between the signal parameters and the search location.

The most significant pulsar candidates from the blind search were automatically refined using the $H$-test statistic \citep{deJager+1989}. This revealed an interesting candidate; however the measured signal-to-noise ratio (S/N) was slightly below the detection threshold for a blind search involving a very high number of trials. Upon manual inspection, clear pulsations were observed in the photon data after April 2010; however the phase of these pulsations was not constant, and exhibited ``wraps'' in which the pulsation phase quickly jumped by one full rotation. These features indicated that the canonical isolated pulsar model used for the blind survey was insufficient, and hence follow-up studies were required to describe the pulsar's rotation over the entire dataset.

\section{Follow-up Analysis}\label{s:timing}
Before carrying out follow-up analyses, we extended the dataset to include photons observed until 2014 October~1 and increased the ROI to 10$\arcdeg$. To speed up the timing procedure computations, we discarded photons with a probability weight below 5\%.

\subsection{Glitch Identification}
\begin{figure*}
\centering
\includegraphics[width=0.93\textwidth]{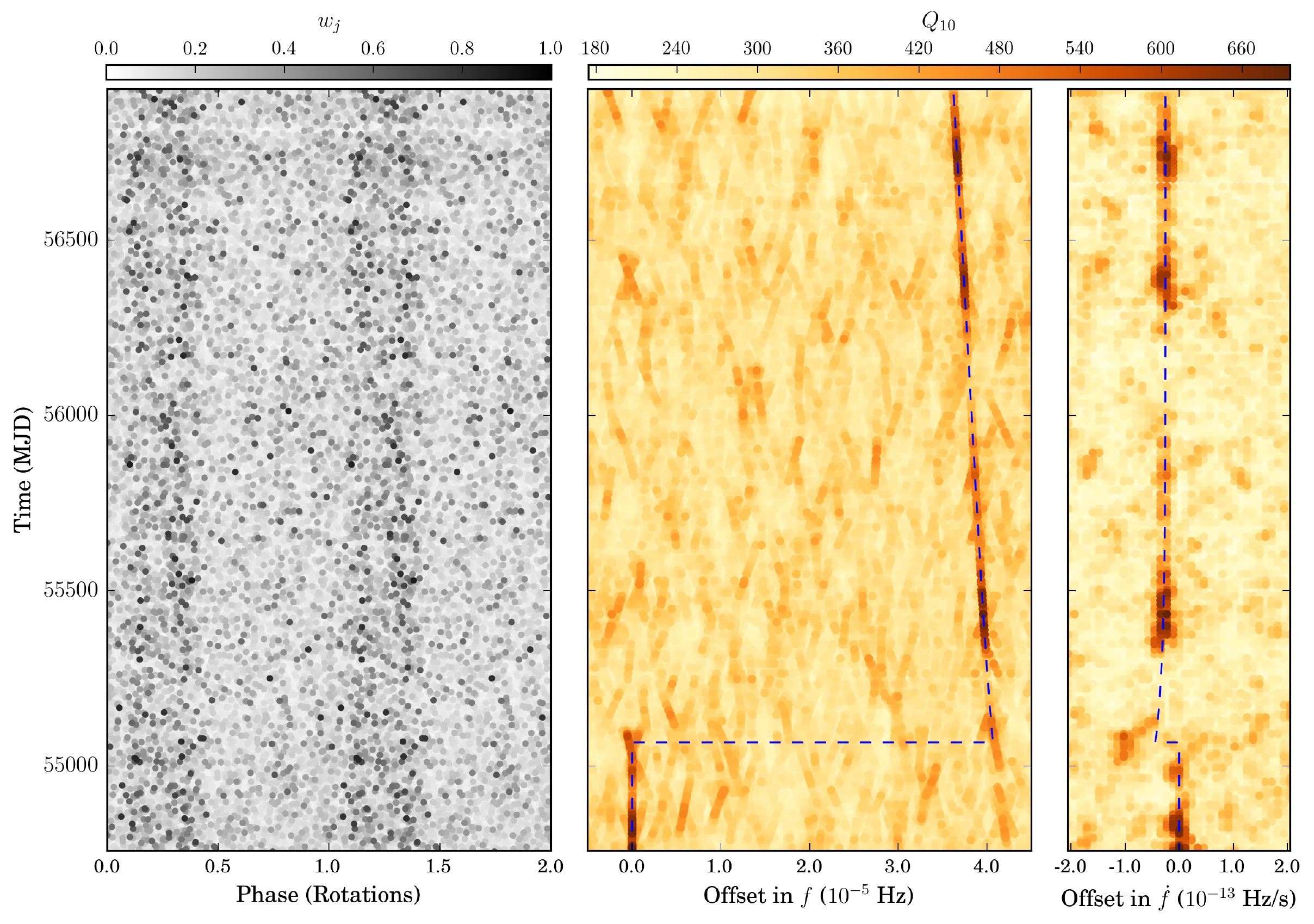}
\caption{Evolution of the PSR~J1906$+$0722 signal including the glitch at MJD 55067. Left: Phase--time diagram where each point represents one photon, with the intensity representing the photon weight. Center and right: The $Q_{10}$-test (shown by the color-bar) calculated over small ranges in $\{f,\dot{f}\}$, centered on the pre-glitch parameters, in overlapping $150$-day segments, and maximized over $\dot{f}$ and $f$ respectively. The dashed line indicates the maximum likelihood timing solution.}
\label{f:spin_evo}
\end{figure*}

Since pulsations were initially only detected during the final 4 years of data, the first step was to identify any glitches within the observation time. To achieve this, we searched the local $\{f,\dot{f}\}$ space around the observed signal, in $150$-day segments with approximately $90\%$ overlap, using the $Q_{M}$-test \citep{Bickel+2008,Methods2014},
\begin{equation}
Q_M = 2M\frac{\sum_{n=1}^M \left|\alpha_n\right|^2 \mathcal{P}_n}{\sum_{n=1}^{M}\left|\alpha_n\right|^2}\,,
\end{equation}
where $\alpha_n$ and $\mathcal{P}_n$ are the Fourier coefficients of the measured pulse profile and the coherent power at the $n$th harmonic respectively. Using the $Q_M$ test method to weight the contributions from each harmonic, as opposed to the commonly used $H$-test, offers a significant sensitivity improvement \citep{Beran1969}, making it particularly useful when analyzing weak pulsar signals. For this step, we included the first 10 Fourier coefficients with appreciable power from a segment of the data in which the signal was reasonably stable.
Using the results of this scan, shown in Figure~\ref{f:spin_evo}, an initial ephemeris was produced for the timing procedure described in the following section.

\subsection{Timing Analysis}
To accurately estimate the pulsar's rotational, glitch and sky location parameters we used a variation of the timing method used by \cite{Ray2011}, based on unbinned likelihood maximization. For all $N$ photons in the dataset, with weights $\{w_j\}$, we assigned a rotational phase $\phi \equiv \phi(t_j,\boldsymbol{u})$, determined by the photon's arrival time, $t_j$ and the set of model parameters, denoted by the vector $\boldsymbol{u}$. For a template pulse profile, $F(\phi)$, the likelihood is
\begin{equation}
  \mathcal{L(\boldsymbol{u})} = \prod_{j=1}^N \left[ w_j\, F\left(\phi(t_j,\boldsymbol{u})\right) + (1 - w_j)\right]\,.
\end{equation}

\begin{figure}
	\centering
	\includegraphics[width=0.95\columnwidth]{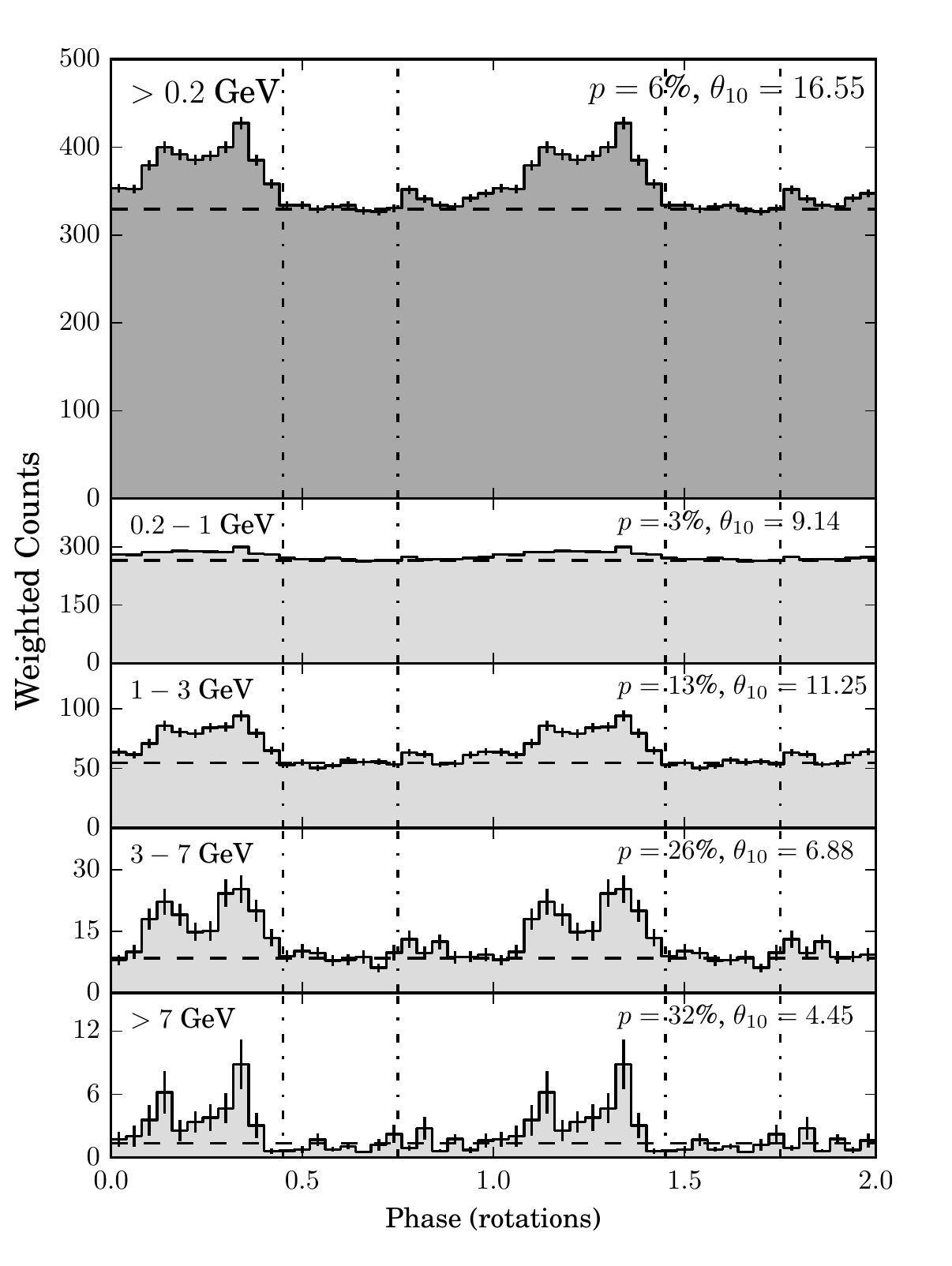}
	\caption{Top panel: weighted pulse profile of PSR~J1906$+$0722 given by the timing 
	solution. Lower panels: weighted pulse profiles in increasing energy bands. In each panel, the estimated background level, calculated from the photon weights \citep{Guillemot2012}, is shown by the dashed line. The dot-dashed lines mark the off-pulse phase interval used in Section \ref{s:off-pulse}. The error bars show $1\sigma$ statistical uncertainties \citep{Pletsch+2012-9pulsars}. The pulsed fraction, $p$, and pulsed S/N, $\theta_{10}$, in each energy band is also shown.}
	\label{f:pulse_profile}
\end{figure}

We first constructed a template pulse profile from the (background subtracted, see Figure \ref{f:pulse_profile}) photons within a sub-section of the data set in which the initial ephemeris was believed to be accurate. When timing PSR~J1906$+$0722 we used a template pulse profile consisting of 3 wrapped Gaussian functions \citep{2PC+2013}, which were fit by maximizing the likelihood within the segment. 

With a template profile at hand, we then estimated the pulsar's parameters (given in Table~\ref{t:params}) by varying them around their initial estimate to maximize the likelihood over the \emph{entire} dataset. The result is a likelihood maximization which is unbinned in both phase (via the template profile) and time. This avoids the need to construct a set of data subsegments for pulse times of arrival (TOA) determination. This is especially beneficial for faint pulsars, which require longer subsegments (and hence fewer TOAs) to ensure the S/N is large enough in each for accurate TOA measurement. Subsequently, using the most likely parameters, the template profile was updated and the process was iterated to maximize the overall likelihood.

\begin{deluxetable}{ll}
	\tablewidth{\columnwidth}
	\tablecaption{\label{t:params} Parameters for PSR~J1906$+$0722}
	\tablecolumns{2}
	\tablehead{
		\colhead{Parameter} &
		\colhead{Value}
	}
	\startdata
	Range of Photon Data (MJD) \dotfill & $54682$--$56931$ \\
	Reference epoch (MJD) \dotfill & $55716$\\
	\cutinhead{Timing Parameters}
	R.A., $\alpha$ (J2000.0) \dotfill & $19^{\rm h}\,06^{\rm m}\,31^{\rm s}.20(1)$ \\
	Decl., $\delta$ (J2000.0) \dotfill & $+07\arcdeg22\arcmin55\farcs8(4)$ \\
	Frequency, $f$ (Hz) \dotfill & $8.9666688432(1)$ \\
	1st frequency derivative, $\dot{f}$, (Hz s$^{-1}$) \dotfill & $-2.884709(2)\times10^{-12}$ \\
	2nd frequency derivative, $\ddot{f}$, (Hz s$^{-2}$) \dotfill & $3.18(1) \times10^{-23}$ \\
	Glitch epoch\tablenotemark{a} (MJD) \dotfill & $55067\substack{+2 \\ -9}$\\
	Permanent $f$ glitch increment\tablenotemark{a}, $\Delta f$ (Hz) \dotfill & $4.033(1)\times10^{-5}$\\
	Perm. $\dot{f}$ glitch increment\tablenotemark{a}, $\Delta \dot{f}$ (Hz s$^{-1}$) \dotfill & $-2.56(3)\times10^{-14}$ \\
	Decaying $f$ glitch increment\tablenotemark{a}, $\Delta f_{\rm d}$ (Hz) \dotfill & $3.64(9)\times10^{-7}$ \\
	Glitch decay time constant\tablenotemark{a}, $\tau_{\rm d}$ (days) \dotfill & $221(12)$\\
	\cutinhead{Spectral Properties}
	Spectral index, $\Gamma$ \dotfill & $1.9\,\pm\,0.1$\\
	Cutoff energy, $E_{\rm c}$ (GeV) \dotfill & $5.5 \, \pm \,  1.2$ \\
	Photon flux\tablenotemark{b}, $F_{100}$ (photons cm$^{-2}$ s$^{-1}$) \dotfill & $(1.1\, \pm \, 0.3) \times 10^{-7}$ \\
	Energy flux\tablenotemark{b}, $G_{100}$ (erg cm$^{-2}$ s$^{-1}$) \dotfill & $(7.3 \, \pm \,   1.3) \times 10^{-11}$ \\
	\cutinhead{Derived Properties}
	Period, $P$ (ms) \dotfill & $111.524136498(1)$ \\
	1st period derivative, $\dot{P}$ (s s$^{-1}$) \dotfill & $3.587895(2)\times10^{-14}$\\
	Weighted $H$-test \dotfill & $731.2$\\
	Characteristic age\tablenotemark{c}, $\tau_{\rm c}$ (kyr) \dotfill & $49.2$ \\
	Spin-down power\tablenotemark{c}, $\dot{E}$ (erg s$^{-1}$) \dotfill & $1.02 \times 10^{36}$ \\
	Surface $B$-field strength\tablenotemark{c}, $B_{\rm S}$ (G) \dotfill & $2.02 \times 10^{12}$\\
	Light-cylinder $B$-field\tablenotemark{c}, $B_{\rm LC}$ (G) \dotfill & $1.34\times10^{4}$ \\
	Heuristic distance\tablenotemark{c}, $d_{\rm h}$ (kpc) \dotfill & $1.91$ \\
	\enddata	
	\tablecomments{Values for timing parameters are the mean values of the marginalized posterior distributions from the timing analysis, with $1\sigma$ uncertainties in the final digits quoted in parentheses.}
	\tablenotetext{a}{Glitch model parameters are defined in \citet{Tempo2}, with the correction noted by \citet{Yu2013+Glitches}.}
	\tablenotetext{b}{Fluxes above 100 MeV, $F_{100}$ and $G_{100}$, were calculated by extrapolation from the $E > 200$ MeV spectrum.}
	\tablenotetext{c}{Derived pulsar properties are defined in \citet{2PC+2013}. The heuristic distance, $d_h = \left(L_{\gamma}^h/4\pi G_{\text{100}}\right)^{1/2}$, is calculated from the heuristic luminosity, $L_{\gamma}^h$, described therein.}
\end{deluxetable}

To explore the multi-dimensional parameter space we used the \texttt{MultiNest} nested sampling algorithm \citep{MultiNest}, which offers high sampling efficiency, and allows posterior distributions to be calculated as a by-product. 

The timing procedure was carried out in two stages: firstly, all timing parameters were allowed to vary. Due to the shortness of the pre-glitch segment, the uncertainties in the glitch parameters dominated those of the remaining timing parameters. We therefore fixed the glitch parameters at their maximum likelihood values, and fit again for the remaining timing parameters.

When timing radio pulsar glitches, \citet{Yu2013+Glitches} noted that unique solutions for glitch epochs could not be found for large glitches occurring during an interval between two radio observations. We observe a similar effect here, although our limiting factor is the photon flux.
When phase folding, a full rotation can be lost/gained if the offset between the model glitch epoch
and the true glitch epoch is more than $1/\Delta f \approx 0.3$ days; however an average of only
$1.4$ weighted photons are observed from the pulsar within this time, making this phase wrap simply
undetectable. We assumed that no phase increment occurred at the glitch, and found that the
posterior distribution for the glitch epoch features several bands, separated by $1/\Delta f$. Due
to the multi-modal shape of the posterior distribution, in Table~\ref{t:params} we report the glitch
epoch that results in the maximum likelihood and the 95\% credible interval.

The inclusion of an additional nearby source in the source model and raising the energy threshold to $200$~MeV when calculating the photon weights for PSR~J1906$+$0722 increased the S/N (see Section~\ref{s:off-pulse}). Therefore the timing analysis was repeated with the updated photon weights, and the results are given in Table~\ref{t:params}.
The time versus rotational phase diagram based on this timing solution is shown in Figure~\ref{f:spin_evo}
and the integrated pulse profile is displayed in Figure~\ref{f:pulse_profile}. Through these refinement and timing procedures, the initial candidate's $Q_{10}$-test S/N \citep{Methods2014}\footnote{The expectation value for $Q_M$ under the null hypothesis, $E_0\left[Q_M\right]$, in \citet{Methods2014} contains an error. S/Ns reported in this work were calculated using the corrected value of $4M$.} was increased from $\theta_{10} = 6.86$ to the highly significant value of $\theta_{10} = 16.55$ given by the final timing solution. 

\subsection{Off-pulse Analysis}\label{s:off-pulse}
Fitting an exponential cutoff model to the spectrum of PSR~J1906$+$0722 revealed a relatively high cutoff energy compared to typical gamma-ray pulsars ($E_{\rm c}$ = 6.5 $\pm$ 0.9 GeV), 
suggesting that the spectrum could be contaminated by the presence of a nearby source 
as was also noted by \citet{Xing2014}.

To investigate this possibility, we analyzed the off-pulse part of the data using photons with energies between 200 MeV and 300 GeV. 
A residual test statistic (TS) map for the off-pulse data (see Figure \ref{f:pulse_profile}) revealed an excess $(0.28 \pm 0.02)^\circ$ away from PSR~J1906$+$0722, at $(\alpha,\delta)$ = ($286.84^\circ$, $7.15^\circ$), with a TS value of $288$.

Modeling this secondary source with a power-law spectrum, we added it to the spectral model for the region, keeping its location fixed from the off-pulse analysis, but leaving its normalization and spectral index free, and analyzed again the full phase interval data. As a result, we found that the log-likelihood value increased slightly, and the new photon weights increased the S/N of the pulsations from $\theta_{10} = 16.38$ to $\theta_{10} = 16.55$. 

The low energy threshold of 200~MeV was chosen to provide improved angular resolution in order to better separate the pulsar emission from that of the new source. When lower energy (100--200~MeV) photons were included in the spectral analysis, the pulsation S/N calculated with the resulting photon weights decreased, suggesting that source confusion at low energies leads to a less reliable source model.  

\begin{figure*}
\centering
\includegraphics[width=0.925\textwidth]{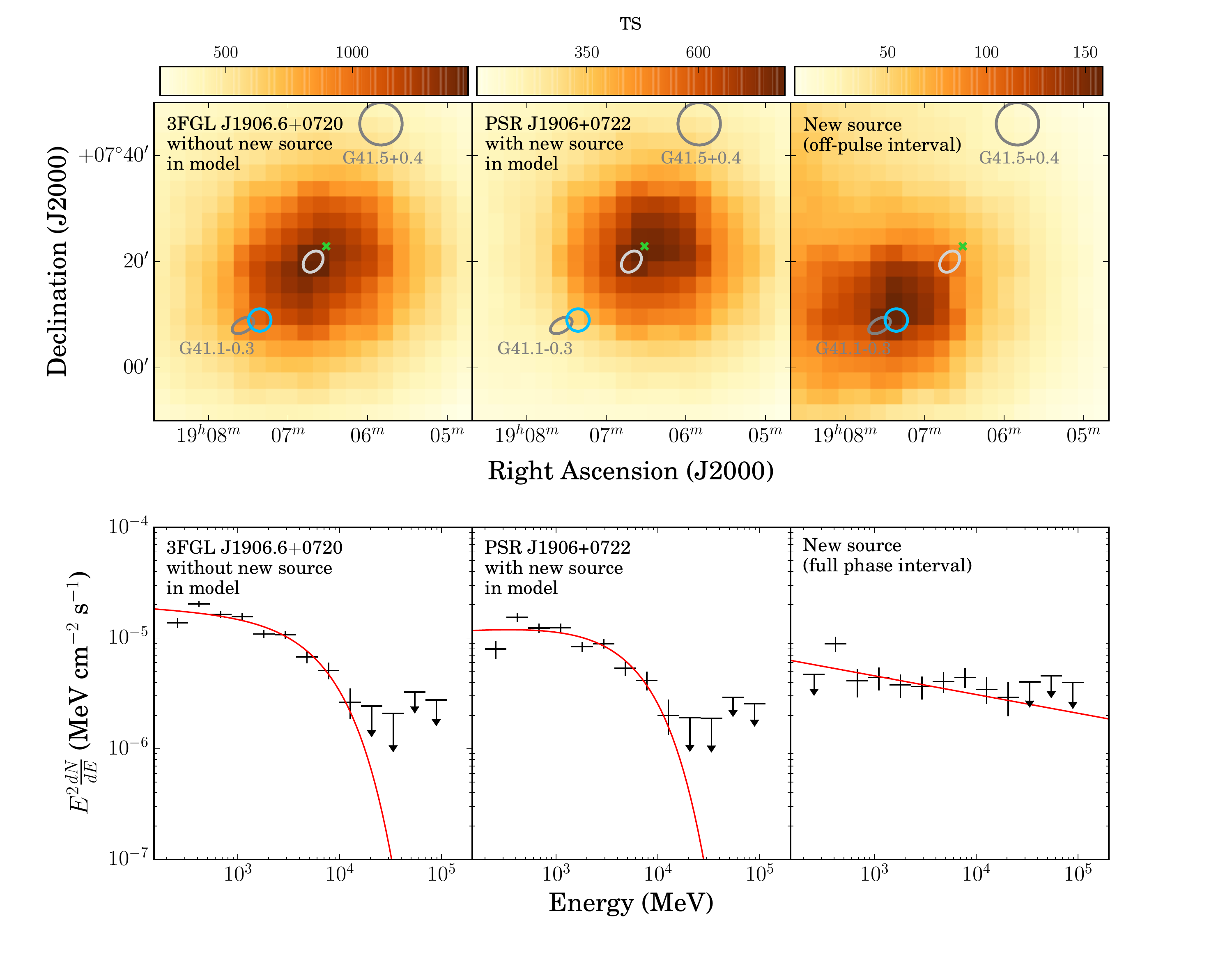}
\caption{Top panels: test statistic (TS) sky-maps of the PSR~J1906$+$0722 region above 200 MeV. Each pixel shows the TS value for a point source located at the pixel position. The cross represents the timing position of PSR~J1906$+$0722. The light ellipse shows the 95\% confidence region of the 3FGL source, the blue ellipse shows the 95\% confidence region of the new secondary source, and the darker ellipses show the approximate extents of nearby SNRs. Bottom panels: Spectral energy distributions for the full-pulse interval. The solid curves present the results of the likelihood analyses of Section \ref{s:off-pulse}.}
\label{f:sed_ts_combine}
\end{figure*}

Figure~\ref{f:sed_ts_combine} shows TS maps and spectral energy distributions 
for PSR~J1906$+$0722 and the new source found in this off-pulse analysis.
The integrated energy flux of the secondary source above 100 MeV is $4.34^{+0.91}_{-0.67}\times10^{-11}$~erg~cm$^{-2}$~s$^{-1}$ with a spectral index of 2.17 $\pm$ 0.07. 

The best-fitting location of the secondary source is very close to the western edge of the supernova remnant (SNR), G41.1-0.3 \citep
[3C 397,][]{Safi-Harb2005+ChandraSNR}. \citet{Jiang2010+MC} observed a molecular cloud interacting with the SNR at this location; it is possible that we are observing gamma-ray emission resulting from this interaction. 

\section{Analysis in Other Wavelengths}

\subsection{Radio and X-ray Observations}\label{s:radio}
In probing for radio emission from PSR~J1906+0722, we carried out a 120-minute follow-up observation with the 
L-band (1.4 GHz) single-pixel receiver mounted on the 100 m Effelsberg Radio Telescope in Germany. 
The gamma-ray-timing ephemeris allowed us to search the data over dispersion measure (DM) only. No
evidence for radio pulsations was found. Assuming a 10\% pulse width, bandwidth $\Delta{F} =
150$\,MHz, telescope gain $G = 1.55$, $n_{p} = 2$ polarization channels, system temperature $T_{\rm
  sys}$ = 24 K, digitization factor $\beta = 1.2$ and a signal-to-noise threshold of $5$, by the
radiometer equation \citep[Equation (A1.22),][p.265]{lorimer2005handbook}, we computed a flux density limit of $\approx 21\,\mu$Jy. While this is below the conventional radio-quiet level of $30\, \mu$Jy \citep{2PC+2013}, we note that the nearby LAT-discovered pulsar PSR~J1907+0602 has been observed in radio observations with a flux density of just $3.4\, \mu$Jy \citep{Abdo2010+J1907}, and would therefore not have been detected in this radio search.

To check for a possible X-ray counterpart, we analyzed a $10$\,ks observation with \textit{Swift}'s
X-ray Telescope \citep{Swift+LATSurvey}. No counterpart source was detected, with an unabsorbed-flux
\mbox{(0.5--10\,keV)} upper limit of $2\times10^{-13}$\,erg\,cm$^{-2}$\,s$^{-1}$ at the pulsar
position.  This limit yields a gamma-ray-to-X-ray flux ratio of $>365$, or an efficiency $L_X/\dot E
\lesssim 8.7\times10^{-5}$ at distance~$d_h$, similar to other gamma-ray pulsars \citep{Marelli2011,
  Pletsch+2012-9pulsars}.

\subsection{Possible SNR Associations}\label{s:snrs}
There are 4 known SNRs lying within $1 \arcdeg$ from the timing position of PSR~J1906$+$0722 \citep{Green+SNRCat}. There is strong evidence that the closest of these, G41.1$-$0.3, is a Type Ia SNR from a Chandrasekhar mass progenitor \citep{Yamaguchi2015+3C397}, making it  unlikely to be the birthplace of a pulsar. Each of the remaining nearby SNRs lies closer to other young pulsars than to PSR~J1906$+$0722 (G41.5$+$0.4 and G42.0$-$0.1 to PSR~J1906$+$0746; G40.5$-$0.5 to PSR~J1907$+$0602), making a physical association between any of these difficult to verify. Kick-velocity requirements based on the pulsar's characteristic age and heuristic distance do not rule out any of these SNRs as the birthplace of the pulsar.

\section{Discussion}\label{s:discussion}
Despite several years of attempts, the identification of 2FGL J1906.5+0720 remained elusive. Now that this source has been identified as PSR~J1906$+$0722, we here investigate potential reasons for the failure of previous searches to detect it. 

Perhaps the most significant source of difficulty in the detection of PSR~J1906$+$0722 was the large positional offset between its 3FGL catalog position and its true position. 
This offset, which could only be accommodated by the new follow-up method outlined in Section~\ref{s:blind_search}, is most likely due to the presence of the secondary source described in Section~\ref{s:off-pulse}.

The close proximity of PSR~J1906$+$0722 to the Galactic plane ($b = 0.03\arcdeg$) likely also hindered its detection, as the large majority of the weighted photons can be attributed to the background. From the pulse profile shown in Figure~\ref{f:pulse_profile}, we estimate that the pulsed fraction of the total weighted photon flux (as defined in \cite{Methods2014}) is as low as $6\%$. This low pulsed fraction leads to a low observable S/N, making detection more challenging.

A further complication for detecting PSR~J1906$+$0722 was the presence of the glitch about one year into the \textit{Fermi} mission. This glitch is among the largest detected from a gamma-ray pulsar in terms of relative magnitude ($\Delta f / f \approx 4.5\times10^{-6}$) \citep{Pletsch2012+J1838}.  
In previous searches using a shorter total observation time, the data segment after the glitch represented a much shorter fraction of the total observation time. 
As the time interval covered by \textit{Fermi}'s observations since 2008 August 1 continues to increase, the existence of a long timespan in which a pulsar's signal is stable becomes ever more likely.
The increase in the weighted photon flux offered by the Pass 8 analysis \citep{Pass8} further increases the observable S/N throughout the observation time, and results in searches that are not only more sensitive overall \citep{Laffon2015+Pass8Pulsars}, but also more robust against glitching or noisy pulsars.

The ability to detect young gamma-ray pulsars in blind searches can be of significant importance to the overall study of energetic pulsars. For example, \cite{Ravi2010} use the observed population of radio-quiet pulsars to investigate the dependence of properties of pulsar emission geometries on the spin-down energy, $\dot{E}$. Since pulsars with a high $\dot{E}$ tend to exhibit timing noise and glitches (which do not typically affect radio searches), they are hard to find in gamma-ray data, where long integration times are required. Advanced search methods that can detect complicated signals such as that from PSR~J1906$+$0722 are therefore crucial for reducing a potential bias against young, energetic and glitching pulsars in the radio-quiet population. As noted by \citet{2PC+2013} and \citet{Caraveo2014}, such pulsars are indeed lacking in the \textit{Fermi} pulsar sample.

\acknowledgements
This work was supported by the Max-Planck-Gesell\-schaft~(MPG), 
as well as by the Deutsche Forschungsgemeinschaft~(DFG) through 
an Emmy Noether research grant PL~710/1-1 (PI: Holger~J.~Pletsch). We are very grateful to all
\textit{Einstein@Home} volunteers who have donated their spare  computing time, especially Connor
Barry of Lafayette, Colorado, USA and Rich Johnson of Hayward, California, USA on whose computers
PSR~J1906$+$0722 was first detected. The \textit{Fermi}-LAT Collaboration acknowledges support for LAT development, operation and data analysis from NASA and DOE (United States), CEA/Irfu and IN2P3/CNRS (France), ASI and INFN (Italy), MEXT, KEK, and JAXA (Japan), and the K.A.~Wallenberg Foundation, the Swedish Research Council and the National Space Board (Sweden). Science analysis support in the operations phase from INAF (Italy) and CNES (France) is also gratefully acknowledged.

\end{document}